\documentstyle[prd,aps,eqsecnum]{revtex}
\newcommand{\bmit}[1]{\mbox{\boldmath $#1$}}

\begin{document}

\draft
\title{Coherent Interaction of a Monochromatic Gravitational Wave 
with both Elastic Bodies and Electromagnetic Circuits}
\author{Enrico Montanari\thanks{Electronic address:
montanari@fe.infn.it} and Pierluigi Fortini}
\address{Department of Physics, University of Ferrara and
INFN Sezione di Ferrara, Via Paradiso 12,
I-44100 Ferrara, Italy}
\maketitle
\begin{abstract}
The interaction of a gravitational wave with a system 
made of an RLC circuit forming one end of a mechanical harmonic 
oscillator is investigated. 
We show that, in some configurations, the coherent 
interaction of the wave with both the mechanical oscillator and the 
RLC circuit gives rise to a mechanical quality factor increase of the 
electromagnetic signal. 
When this system is used as an amplifier of gravitational periodic 
signals in the frequency range $50-1000\,{\rm Hz}$, at ultracryogenic 
temperatures and for sufficiently long integration times (up to 4 
months), a sensitivity of 
$10^{-24}$--$10^{-27}$ on the amplitude of the metric could be achieved 
when thermal noise, shot noise and amplifier back--action are considered.
\end{abstract}
\pacs{PACS number(s): 04.30.Nk, 04.20.-q, 04.80.Nn}

\section{Introduction} 
Gravitational wave detection is a research field which is actively 
under development by many groups all over the world (e.g., 
\cite{amaldi}).
From the theoretical point of view all the devices yet operating or 
under construction could be divided into two classes: the resonant 
detectors 
and the interferometers. In the first case one tries to detect the 
motion of an elastic body due to the gravitational tidal forces 
acting on it. In the second one the gravitational phase shift on an 
electromagnetic wave is exploited.
In both cases the detection is due to {\em only one} kind of direct 
gravitational 
interaction with matter. For instance, in the resonant bar detector 
arrangement, the direct interaction of the gravitational wave with the 
solid body produces mechanical oscillations which are converted into 
an electric signal by a capacitive transducer and then amplified and 
recorded. The bar is the only constituent of the apparatus which 
directly interacts with the gravitational wave. The transducer system 
can not distinguish if the mechanical motion is due to a gravitational 
wave or another source.

In this paper we will study the behaviour of a mechanical oscillator
and an electromagnetic circuit rigidly mounted on one end of it, 
{\em both directly interacting} with a periodic gravitational wave 
(see Fig. 1).
As a first preliminary result, we will find the rest frame of one end of a 
damped oscillator in the field of a gravitational wave, 
generalizing a result contained in~\cite{2}. 
Then we will show that the device described in Fig. 1 is characterized 
by a sort of 
coherent effect: in some configuration the effect is to amplify 
the response of the electromagnetic system by the mechanical quality 
factor of the oscillator. This coherent effect, besides being 
interesting in itself, could be taken into consideration in future 
experiments for almost two reasons: in fact we will show that the 
detection limit of such a device, for periodic sources like pulsars 
(frequency range $50-1000\,{\rm Hz}$), at ultracryogenic temperatures 
and for sufficiently long integration times (up to 4 months) 
seems very interesting, being of the 
order of $10^{-24}$--$10^{-27}$ in the metric perturbation (when 
thermal noise, shot noise and amplifier back--action are taken into 
account); 
moreover, and this is the most important peculiarity, there is a 
decrease of the problem of mechanical insulation, because the 
sensitivity of the circuit can be considered independently of noise 
affecting the mechanical oscillator. This is due to the fact that in this 
arrangement the electromagnetic circuit, in some sense which will be 
clearer afterwards, ``recognizes'' if the 
mechanical oscillator signal is a gravitational one or not.

\section{Gravitational wave interaction with electromagnetic circuits}

Recently it has been shown \cite{1} how, under certain conditions, a set of 
conductors behaves in the field of a gravitational wave. These 
conditions are essentially three. The first one is about the 
electromagnetic field: it is considered in the long--wavelength 
approximation in which the frequency of the electromagnetic 
oscillations $\nu_{em}$ is much smaller than $c/d$ ($d$ is the typical 
linear size of the system and $c$ is the speed of light). The 
second one is about the gravitational field: the spatial distance 
$\lambda_g$ over which the gravitational wave changes appreciably 
must be much greater than $d$. The last condition is about 
the mechanical behaviour of the system: it is freely falling in the 
field of the incoming wave. This implies that it is at rest in 
TT--gauge, reference frame in which the calculation is performed.
Within the framework of these assumptions the equations for a system of 
$N$ conducting non--ferromagnetic one dimensional (wires) or extended 
bodies (capacitors) in the presence of a gravitational wave are derived.
This is made by means of an extension of the lagrangian formalism 
(see, for instance, \cite{l8}), in which the charges on the conducting 
bodies play the r\^ole of generalized coordinates.
We have found that the behaviour of such a system is completely 
described by a lagrangian together with a dissipation function in which 
the coefficients of mutual inductance $L_{\gamma_a\gamma_b}$, the 
inverse of the capacity coefficients $C^{-1}_{\gamma_a\gamma_b}$ and 
the resistances $R_{\gamma_a}(t)$ that characterize the system are 
modified by the gravitational wave (here the subscript $\gamma_a$ 
means the {\em a--th} conductor). This could be summarized by the 
following relations
\begin{eqnarray}
L_{\gamma_a\gamma_b}(t) & = 
&\ ^0\!L_{\gamma_a\gamma_b} + l_{\gamma_a\gamma_b}(t) \nonumber \\
\widehat C_{\gamma_a\gamma_b}(t) & =
&\ ^0\!\widehat C_{\gamma_a\gamma_b} + \widehat c_{\gamma_a\gamma_b}(t) 
\label{pertparam}\\
R_{\gamma_a}(t) & = &\ ^0\!R_{\gamma_a} + r_{\gamma_a}(t) \nonumber 
\end{eqnarray}
where the {\em zero} quantities are the usual flat space--time 
ones, while $l_{\gamma_a\gamma_b}(t)$, 
$\widehat c_{\gamma_a\gamma_b}(t)$ and 
$r_{\gamma_a}(t)$ are the time dependent perturbations
induced by the gravitational waves on the circuit parameters; they 
can be written as
\begin{eqnarray}
l_{\gamma_a\gamma_b}(t) & = & h_{ij}(t) \lambda^{ij}_{\gamma_a\gamma_b}
\nonumber \\
\hat c_{\gamma_a\gamma_b}(t) & = & h_{ij}(t)\chi^{ij}_{\gamma_a\gamma_b}
\label{quadru} \\
r_{\gamma_a}(t) & = & h_{ij}(t)\varrho_{\gamma_a}^{ij}
\nonumber
\end{eqnarray}
where $h_{ij}(t)$ is the metric perturbation which describe the 
gravitational wave in TT--gauge ($g_{\mu\nu}$ is the metric tensor, 
greek indices run from 0 to 3; latin indices from 1 to 3)~\cite{4}
\begin{eqnarray}
g_{\mu\nu} & = & \eta_{\mu\nu} + h{\mu\nu} \nonumber \\
\eta_{\mu\nu} = diag(-1,1,1,1); \qquad & & \qquad
|h_{\mu\nu}| << 1 \label{TTg} \\
h_{0\mu} = 0; \qquad\qquad h = \eta_{\mu\nu} h^{\mu\nu} & = & 0; 
\qquad\qquad h^{ij}_{\ \ ,j} = 0 \nonumber
\end{eqnarray}
and
\begin{equation}
\lambda^{ij}_{\gamma_a\gamma_b}=
{1\over c^2}
\int_{\gamma_a}{}\int_{\gamma_b}{}
{dx^i_{\gamma_a}dx'^j_{\gamma_b}\over
|\bmit x_{\gamma_a}-\bmit x'_{\gamma_b}|}+
{1\over 2c^2}
\int_{\gamma_a}{}\int_{\gamma_b}{\delta_{kl}}
\partial'^i\partial'^j
|\bmit x_{\gamma_a}-\bmit x'_{\gamma_b}|
dx^k_{\gamma_a}dx'^l_{\gamma_b}
\label{lambdaab}
\end{equation}
\begin{equation}
Q^{\gamma_a}\chi^{ij}_{\gamma_a\gamma_b}Q^{\gamma_b}=
{1\over 2}
\int_{\gamma_a}{}\int_{\gamma_b}{}
{dQ^{\gamma_a}\over dA_{\gamma_a}}
{dQ^{\gamma_b}\over dA'_{\gamma_b}}
\partial'^i\partial'^j
|\bmit x_{\gamma_a}-\bmit x'_{\gamma_b}|
dA_{\gamma_a}dA'_{\gamma_b}
\label{chiab}
\end{equation}
\begin{equation}
\varrho_{\gamma_a}^{ij}=
\int_{\gamma_a}{}
{t_{\gamma_a}^it_{\gamma_a}^j\over\sigma_{\gamma_a}}
{dl\over dS_{\gamma_a}}.
\label{rhoa}
\end{equation}
The coefficients $\lambda^{ij}_{\gamma_a\gamma_b}$, 
$\chi^{ij}_{\gamma_a\gamma_b}$ and
$\varrho_{\gamma_a}^{ij}$, defined in 
Eqs.~(\ref{lambdaab})--(\ref{rhoa}), 
are pure geometric quantities which play a similar
r\^ole as the mass-quadrupole tensor
\[
Q^{ij} = \int_V \rho(\bmit x)(x^i x^j -  (1/3) \delta^{ij} x^k x_k)
\ d^3x
\]
in the mechanical interaction between a gravitational wave 
and a solid body. We see therefore that the equations that describe the 
system are parametric ones. This means that gravity {\em is not} 
acting as an electromotive force {\em but} is changing the circuit 
parameters. The effect of the small perturbations of the circuit 
parameters when there is no electrostatic charge distribution is a 
frequency and amplitude modulation (e.g.,~\cite{mac}). 

Let us now consider the circuit described in Fig. 2. It is an RLC 
circuit with two plane and statically charged condensers set at 
right angles.
When a periodic gravitational wave (for the sake of simplicity coming 
from a direction perpendicular to the circuit plane) interacts with 
such a circuit, a current is produced. In this case the effects of 
the parametric interaction given by~(\ref{pertparam}) 
and~(\ref{quadru}) could be summarized by the relations
\begin{eqnarray}
L(t) & = & L [1+\varepsilon_L(t)] \nonumber \\
\frac{1}{C_i(t)} & = & \frac{1}{C_i} [1 - \varepsilon_{C_i}(t)] 
\qquad\qquad i=1,2 \\
R(t) & = & [1+\varepsilon_R(t)] \nonumber
\end{eqnarray} 
where $L$, $C$ and $R$ are the unperturbed inductance, capacitance and 
resistance and the perturbations are very small compared to unity, 
being of the order of the amplitude of the perturbation of the flat 
space--time metric in TT--gauge.
In a reference frame in which the $x$ and $y$ axis are parallel to the 
plates of $C_1$ and $C_2$ respectively then the current flowing is a 
small modulation of the solution of the following equation ($C_1=a C$, 
$C_2 = C$, $\omega_{em}^2=[(a+1)/a] (1/LC)$, $Q_1 = a Q_2$, 
$Q_2=Q_0/(a+1)$)~\cite{1}
\begin{equation}
\ddot Q + 2 \gamma \dot Q + \omega_{em}^2 Q = v_+ h_+(t)
\label{equat}
\end{equation}
where 
\begin{equation}
v_+ = \frac{2 a}{(a+1)^2} \omega_0^2 Q_0
\label{potential}
\end{equation}
and $h_+(t)$ is the ``plus'' TT-gauge polarization amplitude of a 
gravitational wave~\cite{4}.
Eq.~(\ref{equat}) is valid both for the ohmic and the 
superconducting case, provided that $2 \gamma$ is the inverse of the 
damping time of the circuit in the two cases.
The conclusion is that in the case of interaction with circuits having 
a distribution of electrostatic charge 
the most important effect of the gravitational wave is to produce an 
electromotive force having the order of magnitude of the 
electrostatic potential in one condenser times the amplitude of the 
metric perturbation in TT-gauge. If we consider 
Eq.~(\ref{potential}) we can see that it is the analogue of the 
tidal acceleration in the equation of motion of a mechanical 
oscillator. Here the electrostatic charge has taken the place of the 
rest length of the oscillator, and the capacitance that of the inverse 
of the elastic constant of the spring.

In which follows we will study what happens if we rigidly mount with 
no moving parts this particular circuit on one end of a mechanical 
harmonic oscillator.
The key idea is to treat the electromagnetic system in the frame in 
which it is at rest so we can use the results of Ref.~\cite{1}. 

\section{The rest frame}

In the local inertial frame $x^\mu$,
the so--called Fermi Normal 
Coordinates (FNC frame) \cite{4,3}, the interaction with a 
gravitational wave 
of a harmonic oscillator made of two masses $M$, connected with a 
spring of equilibrium length $2L$ in the direction $n^i$ ($L^i=L n^i$), 
in the case in which the wave depends only on time ($L << \lambda_g$), 
is described by the geodesic deviation law (see, for 
instance,~\cite{4}, p.~1023)
\begin{equation}
{\ddot \xi}^i+{  {\dot \xi}^i\over \tau_0 }+\omega_0^2\xi^i= {1\over 2}
 \ {\ddot h}^i_{\ j} x^j\ ,
\label{1}
\end{equation}
where $(L^i + \xi^i)$ are the FNC of one of the two masses,
$\omega_0$ is the proper frequency of the oscillator and $\tau_0$ its
relaxation time. The right--hand side of this equation contains the 
``electric'' components of the Riemann
tensor of the wave in TT coordinates, $R^0_{\ i0j} =
(1/2 c^2) {\ddot h}_{ij}$. In linear approximation the Riemann tensor
is gauge invariant, i.e. it does not depend on the chosen coordinate system.
The solution in Fourier space
[$\xi(\omega)=$ $(2\pi)^{-1/2}\int_{-\infty}^{+
\infty} \xi(t) \exp(-i\omega t) \ dt$], in linear approximation, is 
given by
\begin{equation}
\xi^i(\omega)=
- {1\over 2} L^j h^i_{\ j}(\omega) 
 {\omega^2 \over \omega_0^2-\omega^2 + i\omega/\tau_0 }.
\label{2}
\end{equation}
Within the framework of long wavelength approximation 
($L \ll \lambda_g$), the FNC metric could be written 
as (see, for instance,~\cite{4}, p. 332) 
$\eta_{\mu\nu} + {\cal O}(x^ix^j/\lambda_g^2)$ where $|x^i|\leq L$. 
Therefore up to $(L/\lambda_g)^2$ terms the FNC metric could be safely 
taken as $\eta_{\mu\nu}$ in the region of our interest.

Now we introduce a new coordinate system $y^\mu$ which follows the
motion of the mass. To this aim, we consider the following coordinate 
transformations
\begin{equation}
x^i = y^i - (8 \pi)^{-1/2} y^j 
\int_{-\infty}^{+\infty} d\omega \exp(i\omega t) 
{\omega^2 \over \omega_0^2-\omega^2 + i\omega/\tau_0 } 
h^i_{\ j}(\omega)
\label{3a}
\end{equation}
\begin{equation}
x_0 = y_0 + i (8 \pi)^{-1/2} (y^i y^j/2 c)
\int_{-\infty}^{+\infty} d\omega \exp(i\omega t)
{\omega^3 \over \omega_0^2-\omega^2 + i\omega/\tau_0 } 
h_{ij}(\omega)
\label{3b}
\end{equation}
The new metric (the rest metric) can be written, neglecting 
$(d/\lambda_g)^2$ terms, as [see, for instance,~\cite{l2}, Eq.~(83.6)  
or~\cite{5}, Eq.~(4.3)]
\begin{equation}
\eta_{\mu\nu} + H_{\mu\nu} =  \eta_{\mu\nu} - {1\over \sqrt{2 \pi}} 
\int_{-\infty}^{+\infty} d\omega \exp(i\omega t) 
{\omega^2 \over \omega_0^2-\omega^2 + i\omega/\tau_0 } 
h_{\mu\nu}(\omega)
\label{4}
\end{equation}
In this frame, in the same approximation, the mass under 
consideration is at rest; in fact, 
calling $(L^i + \eta^i)$ its coordinates, then geodesic equation 
becomes 
\begin{equation}
{\ddot \eta}^i+{  {\dot \eta}^i\over \tau_0 }+\omega_0^2\eta^i = 0
\label{5}
\end{equation}
We note that, within the range of the approximations made, 
the rest frame has the same properties~(\ref{TTg}) of a TT--gauge one. 
We further observe that if we consider the limit 
$\omega_0 \rightarrow 0$ and
$\tau_0 \rightarrow \infty$ (free masses) then 
$H_{\mu\nu} \equiv h_{\mu\nu}$ and Eqs.~(\ref{3a}) 
and~(\ref{3b}) become
the transformation laws from the TT--gauge to the FNC to the first 
order in the $L/\lambda_g$ ratio~\cite{4,3}; in other words 
the rest frame becomes 
the TT-gauge (as it must be because it is the rest frame 
for free bodies). 

Let us consider now the case of an electromagnetic circuit rigidly 
mounted on one end of a mechanical harmonic oscillator. In the rest 
frame we are within the framework of the theory summarized in section 
2. We can therefore apply the results of~\cite{1} by simply 
substituting the $h_{ij}$ metric with the rest metric. Therefore the 
rest frame solves the problem of the simultaneous interaction of a 
gravitational wave with a mechanical oscillator {\em and} an 
electromagnetic circuit, being the frame in which only the coupling 
with the latter occurs.

It is to be remarked that we are dealing with 
currents in 
electromagnetic circuits in TT--gauge. In actual experiments 
currents are measured in the laboratory reference frame (FNC). 
However, because of the charge conservation we can write 
\[
I^{FNC}\ dx^0\ =\ I^{TT}\ dy^0
\]
where $I^{FNC}$ is the current measured in the FNC and $I^{TT}$ 
the one measured in a TT--gauge (in this case the rest frame). 
Because of eq.~(\ref{3b}), the relation between currents in 
the two gauges reads
\begin{eqnarray*}
\frac{I^{FNC}}{I^{TT}} &=& 1 -
(8 \pi)^{-1/2} (x^i x^j/2 c^2)
\int_{-\infty}^{+\infty} d\omega \exp(i\omega t)
{\omega^4 \over \omega_0^2-\omega^2 + i\omega/\tau_0} 
h^{ij}(\omega) + \\
&+& i (8 \pi)^{-1/2} \frac{x^i}{c} \frac{dx^i}{dx^0} 
\int_{-\infty}^{+\infty} d\omega \exp(i\omega t)
{\omega^3 \over \omega_0^2-\omega^2 + i\omega/\tau_0 } 
h^{ij}(\omega)  
\end{eqnarray*}
In the long wavelength approximation both first order corrections 
are negligible so that
\[
I^{FNC}\ =\ I^{TT}
\]
(see also \cite{1}).

It is important to observe that the situation envisaged in this 
section {\em is not} equivalent to the case of a 
mechanical oscillator driven by an artificial force. In fact the 
equation of motion of the oscillator in this case is given by
\begin{equation}
{\ddot \xi}^i+{  {\dot \xi}^i\over \tau_0 }+\omega_0^2\xi^i= f^i(t)
\label{ref1}
\end{equation}
where $f^i(t)$ are three functions of time. If we now 
introduce a rest frame, that is to say a frame comoving with the mass 
element of the mechanical oscillator over which is mounted the 
electromagnetic device, and calculate the metric tensor in this frame,
we find that
the metric is unchanged. 
In fact, if $\xi^i = \zeta^i(t)$ is the solution of Eq.~(\ref{ref1}), 
then the coordinate system $\hat y^\mu$ of the rest frame 
is given by
\begin{equation}
x^i = \hat y^i + \zeta^i(t);\qquad\qquad x^0 = \hat y^0
\label{ref2}
\end{equation}
The new metric will then be given by (it has to be remembered that in 
the long wavelength approximation the FNC perturbation of the metric 
is negligible)
\begin{equation}
\hat h_{ij} = 0; \qquad \hat h_{0i} = \zeta_{i,0};
\qquad \hat h_{00} = 0
\label{ref3}
\end{equation}
In the same approximation as before $\zeta_{i,0}$ could be neglected; 
therefore we are led to the conclusion that
\begin{equation}
\hat h_{\mu\nu} = 0
\end{equation}
This result shows the difference between an external driving 
force on the mechanical oscillator and a gravitational 
wave, difference which lies in the tidal nature of the latter.

\section{Monocromatic Waves and Coherent Interaction}

Let us consider now the case of a monochromatic gravitational wave of 
angular frequency $\omega_g$. The coordinate system $y^\mu$ can be 
now written as [$r=(\omega_g/\omega_0)$; 
${\cal Q}_m = (\omega_0\tau_0)/2$ is the mechanical quality factor]:
\begin{eqnarray}
x^i &=& y^i + A(r, {\cal Q}_m) y^j h^i_{\ j} + 
B(r,{\cal Q}_m) y^j \dot h^i_{\ j}/\omega_g \label{6} \\
x_0 &=& y_0 - (y^i y^j/2 c) \left [A(r, {\cal Q}_m) \dot h_{ij} -
\omega_g B(r,{\cal Q}_m) h_{ij}\right ] \nonumber
\end{eqnarray}
while the rest metric takes the form
\begin{equation}
H_{ij} = 2 \left [A(r, {\cal Q}_m) h_{ij} + 
\frac{B(r,{\cal Q}_m)}{\omega_g}\dot h_{ij}\right ]
\label{7}
\end{equation}
where
\begin{eqnarray}
A(r, {\cal Q}_m) &=& -{1\over 2} 
{r^2 (1 - r^2)\over (1 - r^2)^2 + r^2/4{\cal Q}_m^2} 
\label{8a} \\
B(r,{\cal Q}_m) &=& {1\over 4 {\cal Q}_m}
{r^3\over (1 - r^2)^2 + r^2/4{\cal Q}_m^2}
\label{8b}
\end{eqnarray}
In the resonant case, $\omega_g=\omega_0$, $A(1, {\cal Q}_m) = 0$ 
while $B(1,{\cal Q}_m) = {\cal Q}_m$ so that the rest metric at 
mechanical resonance becomes:
\begin{equation}
H_{ij}^{(mr)} = {2 {\cal Q}_m\over \omega_g} \dot h_{ij}
\label{9}
\end{equation}
Therefore the interaction of the gravitational wave at mechanical 
resonance produces, on the electromagnetic 
circuit, a sort of effective metric perturbation 
larger by a factor $2{\cal Q}_m$ and $\pi/2$ phase shifted.

\section{Independence of circuit sensitivity on noise affecting the 
mechanical oscillator}

The amplification mechanism described in previous Section concerns
only gravitational signals that coherently 
interact with both the mechanical oscillator and the 
electromagnetic device. In fact, if the mechanical 
oscillator is accelerated, the order of magnitude of the inertial 
electromotive force induced on the electromagnetic system is 
\begin{equation}
{\cal E}_{in} \approx {m_e\over e} a d
\label{10}
\end{equation}
where $m_e$ is the electronic mass $e$ its charge and $a$ the 
magnitude of the acceleration and $d$ the linear size of the circuit.. 
If the energy of the motion (for instance thermal 
noise) is comparable with the one due to the gravitational wave 
interaction then $a=\omega_g^2 L h$ where $h$ is the order of 
magnitude of the gravitational wave amplitude.

Let us consider now the circuit of Fig. 2. The order of magnitude of 
the electromotive force induced by a gravitational wave with $h$ 
amplitude is [see Eq.~(\ref{potential}) or~\cite{1}]
\begin{equation}
{\cal V} \approx V_0 h
\label{11}
\end{equation}
where $V_0$ is the order of magnitude of the electrostatic potential 
in the condensers.
The ratio between these two electromotive forces is
\begin{equation}
{{\cal E}_{in}\over {\cal V}} \approx 
{m_e\over e} {\omega_g^2\over V_0} L d \simeq 
2 \times 10^{-10} {\nu_g^2\over V_0}
\label{12}
\end{equation}
where we have put $L d = 1\,{\rm m}^2$, $\nu_g$ is expressed in Hz and 
$V_0$ in Volts. Even setting $V_0 = 1\,{\rm V}$ and
$\nu_g = 10^3\,{\rm Hz}$ we obtain $= 10^{-4}$ for the ratio.
In conclusion, in order to calculate the sensitivity of 
the device it suffices to consider noise only on the 
electromagnetic circuit. Hence, at mechanical resonance, the 
sensitivity in $h$ is increased by the mechanical quality factor 
${\cal Q}_m$ (see Eq.~(\ref{9})].

\section{Discussion}

In this Section noise is evaluated in order to obtain the sensitivity of 
such a device.
Taking into account discussion made in Sec. V, only noise
on the electromagnetic circuit is considered. We take into account
thermal mechanical noise, thermal electric noise, 
back action of the amplifier and shot noise of the 
difference of the leakage current of the condensers.

Mechanical thermal noise describes the relative variation of length 
of inductance (or distance between plates of condensers) caused by 
thermal noise. 
In the free falling mechanical approximation (see~\cite{1}), the
amplitude of the gravitational wave having a signal to mechanical noise 
ratio equal to 1 after an observation time $t_{obs}$, is 
[see Eq.~(\ref{9})]:
\begin{equation}
h_{mtn} = \frac{1}{{\cal Q}_m}\,\sqrt{\frac{4\,k_B\,T}
{M\,l^2\,\tau\,\omega_g^4\,t_{obs}}}
\label{hmtn}
\end{equation}
where $k_B$ is the Boltzman constant, $T$ the temperature of 
mechanical degree of freedom, and $M$, $l$ and $\tau$ are the mass, 
length and damping time of mechanical oscillators describing 
condensers or inductance from the mechanical point of view.  
The analogue for the electric thermal noise at electric resonance 
($\omega_g=\omega_{em}$) is [see~\cite{1}, Eq.~(6.44)]:
\begin{equation}
h_{etn} = \frac{1}{{\cal Q}_m V_0}\,
\sqrt{\frac{k_B T}{{\cal Q}_{em} C \omega_g t_{obs}}}
\label{13}
\end{equation}
where ${\cal Q}_{em}$ is the electric quality factor of the circuit, 
and $C=C_1=C_2$.

Another noise source to take into account for is the shot noise of the 
difference, $\Delta I$, of the leakage current of $C_1$ and $C_2$. The 
gravitational signal with a signal to leakage noise ratio equal to $1$ 
after an observation time $t_{obs}$ is [see~\cite{mours} 
and~\cite{1}, Eqs.~(6.15) and~(6.26)]
\begin{equation}
h_{ln} = \frac{1}{4\,{\cal Q}_m\,\omega_g\,V_0\,C\,{\cal Q}_{em}}
\sqrt{\frac{2\,e\,\Delta{I}}{t_{obs}}} 
\label{leakage}
\end{equation}
where $e$ is the absolute value of the electronic charge and we can 
set $\Delta I=10\,pA$~\cite{mours}. 

Once the bar and the electromagnetic circuit are isolated from the 
outside world by means of mechanical filters and Faraday cages
a further noise source is the back--action of the 
amplifier. However, at cryogenic temperature and for the best 
amplifiers, this is at most of the same order of magnitude of thermal 
noise.
Therefore, the amplitude of the gravitational radiation having a signal 
to noise ratio equal to $1$ is 
given by 
\begin{equation}
h_n = \sqrt{2\,\left (h_{mtn}^2 + h_{etn}^2\right ) + h_{ln}^2}
\label{totnoise}
\end{equation}

In order to obtain the order of magnitude of $h_n$, the parameters 
which enter in Eqs.~(\ref{hmtn})--(\ref{totnoise}) had to be set.
Attention must be paid to the fact that weight and size 
of the electric circuit are constrained by the mechanical harmonic 
oscillator to one end of which the circuit is rigidly mounted.
A possible choice is: ${\cal Q}_m = 10^6$, $M = 1\,{\rm Kg}$,
$\tau = 1\,{\rm sec}$, $l=10^{-2}\,{\rm m}$, 
$t_{obs} = 10^7\,{\rm sec} \sim 4\,{\rm months}$, 
$V_0 = 10\,{\rm kV}$, $C = 1\,\mu{\rm F}$, 
${\cal Q}_{em} = 10^3$ in the ohmic case and $10^6$ for superconducting 
circuit. Moreover we assume to operate at cryogenic ($4\,{\rm K}$) and
ultracryogenic ($40\,{\rm mK}$) temperatures and for
gravitational wave frequencies of $60$ and $600\,{\rm Hz}$. The reason 
of this last choice is twofold: first, the possibility to realize 
resonant mechanical detectors operating at these 
frequencies~\cite{6,amaldi}; second, the presence of possible almost
monochromatic sources such as crab 
pulsar~\cite{suz95} or millisecond pulsars~\cite{pulsar}.
In this range it is found that the leakage noise is negligibly small.
Moreover for superconducting circuit, mechanical thermal noise is 
always the biggest noise source, while in the ohmic case, at the higher 
gravitational frequency, electric thermal noise becomes more relevant 
than mechanical one. The results can be found in Tab.~\ref{tab1}.

In order to make comparisons with the predicted sensitivities of 
ground based interferometric detectors such as VIRGO and LIGO 
(e.g.,~\cite{amaldi}), 
for gravitational signals from pulsars, we 
introduce the so called characteristic amplitude $h_c$ defined in
Ref.~\cite{thorne}, Eq.~(50); setting $A_+$ the ``plus'' TT--gauge 
amplitude polarization we have:
\begin{equation}
h_c = \sqrt{\frac{2}{15}}\cdot \frac{A_+}{1 + \cos^2{i}}
\label{char}
\end{equation}
where $i$ is the angle between the rotation axis and the line of sight 
to the source. Therefore the amplitude $h_c$ of the weakest source 
detectable with a signal to noise ratio equal to $3$ in 
$4\,{\rm months} = 10^7\,{\rm sec}$ integration time, with known 
frequency, is therefore given by
\begin{equation}
h_c = \sqrt{\frac{6}{5}}\cdot \frac{h_n}{1+\cos^2{i}}
\label{hcdet}
\end{equation}
The results for ohmic and superconducting circuit are given in 
Tabs.~\ref{tab2} and~\ref{tab3}.
The incertitude takes into account 
for different possible orientations of the rotation axis.

A comparison with LIGO sensitivities (see, for instance,~\cite{brady}) 
shows that, 
near $60\,{\rm Hz}$, the expected sensitivity is of the order of 
$h_c \simeq 10^{-25}$ for initial detector and
$h_c \simeq 2 \times 10^{-27}$ for advanced detector, while near
$600\,{\rm Hz}$ the expected sensitivities are $3 \times 10^{-25}$ and
$10^{-26}$ for initial and advanced detector respectively.

Summarizing, at ultracryogenic temperatures, a device such as that
described in this paper 
could have the same sensitivity as a ground based interferometer of 
the first generation near $60\,{\rm Hz}$, while for higher frequencies 
the sensitivity seems to be comparable also with that of an advanced 
interferometer.
These preliminary results are encouraging and show that a device 
exploiting coherent interaction amplification could be interesting in 
the field of gravitational wave detection.

\section{Conclusions}
We have found the rest frame for one end of a damped harmonic 
oscillator in the field of a gravitational wave. 
This rest frame has been used to obtain the current flowing in an RLC 
circuit with an electrostatic charge, rigidly mounted on that end and 
interacting itself with the wave. 
The result on the circuit behaviour could be described as a 
coherent effect 
between two electromotive forces always in phase: the inertial one
(produced by the elastic motion of the harmonic oscillator in the 
field of the gravitational wave) and the one induced directly on the 
circuit by the wave.
Together with this it must be noted that, not having a definite 
phase, inertial electromotive forces due to noise bar vibrations do 
not couple coherently with the circuit.
Therefore an RLC circuit mounted on one end of 
a mechanical bar detector operating on the same frequency could be an 
interesting amplifier
of gravitational 
periodic signals, as for instance those produced by pulsars and in 
particular millisecond pulsars~\cite{pulsar}. 
This new way of amplify signals
could 
reduce
the problems of the mechanical insulation of the bar, because 
noise bar vibrations do not couple with the circuit with the same 
strength as the case of vibration of gravitational origin.

\acknowledgments

The authors would like to thank S. P. Vyatchanin, V. Ferrari, 
A. Giazotto, G. Sch\"afer, M. Calura and G. Di Domenico for useful 
discussions. 
One of the authors (E. M.) would like to thank A. Drago, V. Guidi,
G. Pareschi and D. Etro for their encouragement.

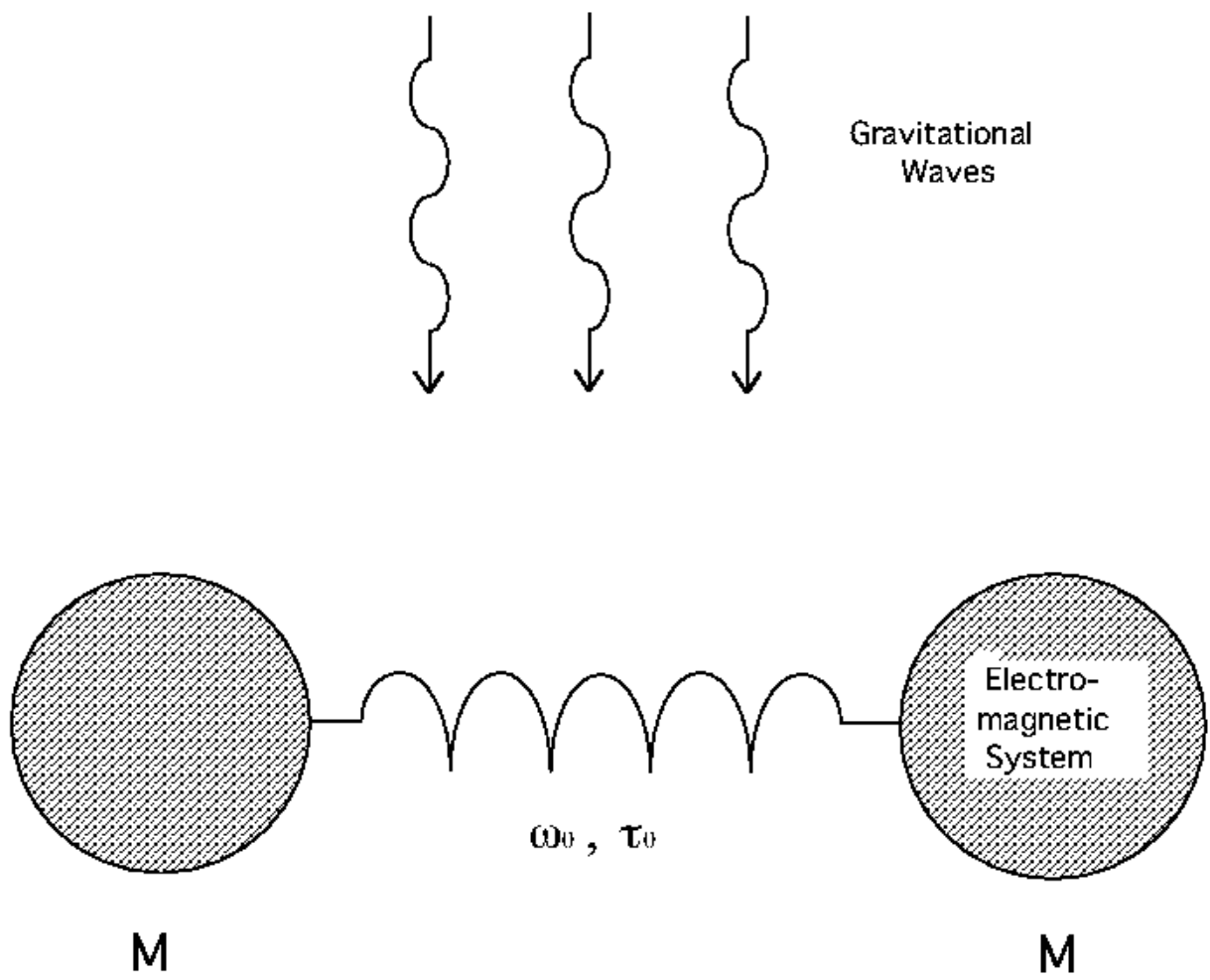
\begin{figure}
\caption{An electromagnetic system is rigidly mounted on 
a side of a mechanical harmonic oscillator with proper frequency 
$\omega_0$ and relaxation time $\tau_0$. Both of them interact with 
a gravitational wave.}
\end{figure}

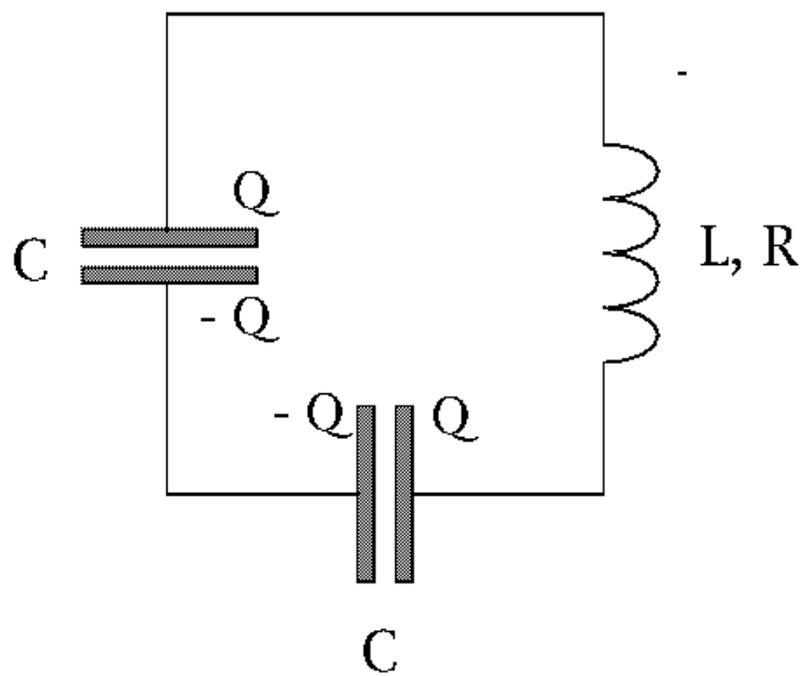
\begin{figure}
\caption{RLC circuit. A static charge ($Q_1\cdot Q_2 > 0$) 
is distributed on the plates of the condensers $C_1$ and $C_2$.}
\end{figure}


\begin{table}
\caption{Gravitational amplitudes with signal to noise ratio equal to 
$1$ in $10^7\,{\rm sec} \simeq 4\,{\rm months}$ integration time for 
different operating temperatures and gravitational frequencies.
$h_{mtn}$, $h_{etn}$ and $h_{ln}$ refer to mechanical thermal noise, 
electric thermal noise and leakage noise, respectively.\\
\label{tab1}}
\begin{tabular}{|c|c|c|c|c|} 
\multicolumn{5}{|c|}{Ohmic circuit} \\ 
\hline
temperature (K) & gravitational frequency (Hz) & 
$h_{mtn}$ & $h_{etn}$ & $h_{ln}$ \\ 
\hline
$4$ & $60$ & $3.3\cdot 10^{-24}$ & $3.8\cdot 10^{-25}$ &
$3.8\cdot 10^{-29}$ \\ 
\hline 
$4$ & $600$ & $3.3\cdot 10^{-26}$ & $1.2\cdot 10^{-25}$ &
$3.8\cdot 10^{-30}$ \\
\hline   
$0.04$ & $60$ & $3.3\cdot 10^{-25}$ & $3.8\cdot 10^{-26}$ &
$3.8\cdot 10^{-29}$ \\
\hline
$0.04$ & $600$ & $3.3\cdot 10^{-27}$ & $1.2\cdot 10^{-26}$ &
$3.8\cdot 10^{-30}$ \\
\hline
\multicolumn{5}{|c|}{Superconducting circuit} \\ 
\hline
temperature (K) & gravitational frequency (Hz) & 
$h_{mtn}$ & $h_{etn}$ & $h_{ln}$ \\ 
\hline
$4$ & $60$ & $3.3\cdot 10^{-24}$ & $1.2\cdot 10^{-26}$ &
$3.8\cdot 10^{-32}$ \\
\hline
$4$ & $600$ & $3.3\cdot 10^{-26}$ & $3.8\cdot 10^{-27}$ &
$3.8\cdot 10^{-33}$ \\
\hline
$0.04$ & $60$ & $3.3\cdot 10^{-25}$ & $1.2\cdot 10^{-27}$ &
$3.8\cdot 10^{-32}$ \\
\hline
$0.04$ & $600$ & $3.3\cdot 10^{-27}$ & $3.8\cdot 10^{-28}$ &
$3.8\cdot 10^{-33}$ \\
\hline
\end{tabular}
\end{table}

\begin{table}
\caption{Characteristic amplitude $h_c$ of the weakest monochromatic 
signal from a pulsar detectable with a signal to noise ratio equal to
3 in $4$ months integration time with ohmic circuit. The incertitude 
takes into account for different possible orientations of the rotation 
axis of the pulsar.\\
\label{tab2}}
\begin{tabular}{|c|c|c|} 
  & $\nu_g = 60$ Hz & $\nu_g = 600$ Hz \\
\hline 
$4$ K & $(2.6-5.2) \cdot 10^{-24}$ & $(1.0-1.9) \cdot 10^{-25}$ \\
\hline
$40$ mK & $(2.6-5.2) \cdot 10^{-25}$ & $(1.0-1.9) \cdot 10^{-26}$ \\
\hline
\end{tabular}
\end{table}

\begin{table}
\caption{The same as Tab.~\ref{tab2}, but for superconducting circuit.\\
\label{tab3}}
\begin{tabular}{|c|c|c|} 
  & $\nu_g = 60$ Hz & $\nu_g = 600$ Hz \\
\hline 
$4$ K & $(2.6-5.1) \cdot 10^{-24}$ & $(2.6-5.2) \cdot 10^{-26}$ \\
\hline
$40$ mK & $(2.6-5.1) \cdot 10^{-25}$ & $(2.6-5.2) \cdot 10^{-27}$ \\
\hline
\end{tabular}
\end{table}

\end{document}